# Shape resonance at a Lifshitz transition for high temperature superconductivity in multiband superconductors

Davide Innocenti

*Advanced Light Source, Lawrence Berkeley National Laboratory, One Cyclotron Road, Berkeley, California 94720, USA, and Dipartimento di Ingegneria Meccanica, Università di Roma Tor Vergata, I-00133 Roma, Italy*

Antonio Valletta

*Institute for Microelectronics and Microsystems, IMM CNR, Via del Fosso del Cavaliere 100, 00133 Roma, Italy*

Antonio Bianconi

*Physics Department, Sapienza University of Rome, P.le Aldo Moro 2, 00185 Roma, Italy*

*abstract*

We discuss the shape resonance in the superconducting gaps of a two band superconductor by tuning the chemical potential at a Lifshitz transition for Femi surface neck collapsing and for spot appearing. The high temperature superconducting scenario for complex matter shows the coexistence of a *first BCS condensate* made of Cooper pairs in the first band and a second *boson-like condensate* made of bosons like bipolarons, in the second band where the chemical potential is tuned near a Lifshitz transition. The interband coupling controls the shape resonance in the *pair exchange between the two condensates*. We discuss the particular BCS-Bose crossover that occurs at the shape resonance tuning the Lifshitz parameter (the energy difference between the chemical potential and the Lifshitz topological transition) like tuning the external magnetic field for the Feshbach resonances in ultracold gases. This superconducting phase provides a particular case of topological superconductivity with multiple condensates of different winding numbers.

1. **Introduction**

Standard BCS superconductivity is the dynamical ordered phase of simple Fermi systems at T=0 on the contrary high temperature superconductivity (HTS) [1] occurs in complex systems [2,3]. HTS is considered the prototype for unknown macroscopic quantum coherent phases that resist to the de-coherence attacks of temperature. There is a growing interest on these phases of complex matter since quantum collective phenomena could be the driving mechanisms for the living systems.[4,5,6]

The complexity of cuprate superconductors making these materials a type of soft dynamical matter has emerged slowly in these last 24 years of experimental research :

a) the charge carriers in the conduction bands have multiple orbital characters: itinerant holes in the oxygen orbital of the $CuO_2$ plane [7] and holes in the Cu 3d orbital;

b) the critical temperature increases with the appearing of a second Fermi surface: the chain band in $YBa_2Cu_3O_{6+y}$ [8] known since 1988, and the second band recently observed in the optimally doped trilayer cuprate Bi2223.[9]

c) a spatial phase separation into two metallic phases made of distorted and undistorted stripes at the optimum doping regime [10] and two electronic states near the Fermi level.[11,12]

d) the presence of a quantum critical point [13,14] in the two variables phase diagram: doping and misfit strain, where the characteristic feature of complexity: scale invariance appears.[15]

Here we report additional support for the 1993 proposed paradigm [16] for understanding HTS, that allows the material design of new HTS,[17,18] based on the fact that:

1) HTS appears in heterostructures at atomic limit, like superlattices of quantum wires, superlattices of quantum stripes and superlattices of quantum wells.

2) The electronic structure of HTS is made of multiple non-hybridized electronic components forming multiple Fermi surfaces. In cuprates a first itinerant electron gas with pure $b_1$ orbital symmetry forming a Fermi pocket at $(\pi,\pi)$ coexists with a polaronic gas with mixed $b_1+a_1$ orbital character forming a Fermi pocket at $(\pi,0)$ [16,19-25]

3) The HTS phase appears by tuning the chemical potential at a shape resonance in the superconducting gaps [26-31] in multiband superconductors.[32]

Several practical realizations of the material design protocol [17,18] for high temperature superconductivity tuned at a shape resonance have been reported for superlattices of superconducting layers made of:

i) boron honeycomb layers separated by Al, Mg or Sc layers, (called diborides) [33-38]

ii) honeycomb graphene layers intercalated by different spacer layers (called intercalated graphite) [39]

iii) iron fcc layers intercalated by many different types of spacers (called pnictides or iron-based superconductors) [40-47],

iv) superlattices of carbon nanotubes [48]

In fact all these systems show similar features [49-52] being multilayers near a Lifshitz transition. In this 2.5 electronic phase transition [53-60] the system is in the verge of a first order phase transition, with the possible appearance of a tricritical point. Phase separation has been observed experimentally and several theoretical models have been proposed [61-64] involving lattice, charge and spin fluctuations near a lattice instability for structural phase transitions [41-43]. The complex phase of matter favoring HTS has been called *superstripes* [64,43] since it appears in low-dimensional systems having relevant similarity with multiscale phase separation in living matter [3].

Multiband superconductivity is now emerging as a fundamental common feature in the Fermiology of cuprates [65-75], diborides [36] and pnictides [45-47] where the chemical potential is tuned in the proximity of a quantum critical Lifshitz transition. There is now growing agreement that the critical temperature appears by a fine tuning of the chemical potential near a Lifshitz transition [76-79] where the superconductivity involve the change of the symmetry of the Fermi surface and of the superconducting condensate.[80] A similar scenario is possible in ultracold gases [81] since the shape resonance is a fundamental concept for many body systems in nuclear, atomic, molecular and condensed matter.[82]

While the consensus on multiband superconductivity in the clean limit has been rapidly accepted for diborides and pnictides [44-47] only recently [65-75] the Fermi surface of high-temperature cuprate superconductors has been mapped to high resolution by the quantum oscillations over a wide range in magnetic field and a large consensus has been reached on the two band superconductivity in cuprates. The Fermi surface comprises multiple pockets, as revealed by the additional distinct quantum oscillation frequencies and harmonics and angle-resolved measurements made of two similarly sized pockets with greatly contrasting degrees of interlayer corrugation. The first hole pocket that is almost ideally two-dimensional in form (exhibiting negligible interlayer corrugation) at the 'nodal' point will correspond to a first subband in our model. The newly resolved weaker adjacent spectral features originating from a deeply corrugated pocket at the

'antinodal' point within the Brillouin zone correspond to electrons expected to be responsible for the negative Hall coefficient.

Moreover recently the multilayer band splitting in two bands (OP and IP) has been observed in the optimally doped trilayer cuprate Bi2223 by angle-resolved photoemission spectroscopy. The OP band is overdoped with a large d-wave gap around the node of 43 meV while the IP is underdoped with an even large gap of 60 meV. [9] These energy gaps are much larger than those for the same doping level of the double-layer cuprates, which leads to the large Tc in Bi2223.

In the multiband theory the role of interband pairing as the key term for rising the critical temperature of the superconducting phase in all known high temperature superconductors is getting a strong experimental support. Moreover there is increasing evidence that the critical temperature reaches a maximum in the proximity of a Lifshitz transition associated with the change of the Fermi surface topology of one of the bands. This scenario is called the "shape resonance" scenario [26-31] that is related with the Majorana configuration interaction between open and closed channels [82] for quasi-stationary states embedded in the continuum and topological superconductors involving condensates with different winding numbers.[80]

In this work we describe the complex physics of two components superfluid condensates with two order parameters: the first is a standard BCS condensate and the second is a bosonic condensate near a band threshold. The Fermi level is close to the band edge in the second band. In a single band this is the scenario of BCS condensation to Bose-Einstein-Condesation (BCS-BEC) crossover discussed by Nozieres [83,84] but he did not consider two-gap picture as we have done. Of course, the two-gap you we considering is bringing a new dimension and importance to the problem that is possible to solve considering the renormalization of the chemical potential in the superconducting phase. In fact at zero temperature for a system of 3D fermions interacting via a contact pairing interaction can be described with continuity by the BCS (mean-field) equations. Indeed, as shown by Leggett,[85] the ground-state BCS wave-function corresponds to an ensemble of overlapping Cooper pairs at weak coupling (BCS regime) and evolves to molecular (non-overlapping) pairs with bosonic character as the pairing strength increases (BEC regime). The crucial point is that the BCS equation for the superconducting gap has to be coupled to the equation that fixes the fermions density: with increasing coupling (or decreasing density), the chemical potential results strongly renormalized with respect to the Fermi energy of the non interacting system.[52] The molecular binding energy of the corresponding two-body problem in the vacuum is very difficult to treat in a single band system since all charge carriers near the band edge condense, the correlation length is of the order or particle distance and the condensate evolves from a Bose to a BCS regime. We show that BCS-BEC regime can be described in the two-bands scenario if the multi-gap superconducting phase is treated theoretically avoiding all standard BCS approximations: 1) high Fermi energy; 2) anisotropic interactions; 3) negligible shift of the chemical potential going from the normal to the superconducting phase; 4) constant density of states; and 5) dirty limit. We provide the case of a chemical potential near a Lifshitz transition of the first type "New spot appearing" and of the second type "Neck collapsing" in second band. We show that below and at the "New spot appearing" a mixed Bose-Fermi regime appears and the Shape resonance shows an anti-resonance regime with a very low $T_c$, on the contrary in the proximity of the "Neck collapsing" or "disruption of a neck" the maximum critical temperature is reached.

## 2. The Lifshitz transition of the type "Neck collapsing"

The shape resonance in the superconducting gaps occurs by tuning the chemical potential at a Lifshitz transition in one of the bands of a metallic system made of non-hybridized bands crossing the Fermi level. A quantum critical point appears in the proximity of a Lifshitz transition with typical quantum criticalities and possible quantum tricritical behavior in itinerant electron systems. [58-60] There are two types of Lifshtz transitions: type I, the "appearance of a new detached Fermi surface region" or appearing or disappearing of a new Fermi surface (FS) spot and type II, disruption of the neck or a neck-collapsing-type Lifshitz transition that can be induced by orbital symmetry breaking in lightly hole doped bands. In Fig. 1 we show that a new 3D Fermi surface (FS) opens when the chemical potential crosses the energy edge $E_{edge}$, and the electron gas in the metallic phase undergoes an electronic topological transition (ETT). When the chemical potential is beyond the band edge in a anisotropic system at a higher energy threshold, $E_{3D-2D}$ the electronic structure undergoes a second ETT, the 3D-2D ETT where one FS changes topology from 3D to 2D (i.e., from "spherical" to "cylindrical") or vice versa, called also "the opening or closing of a neck in a tubular FS" or "Neck collapsing".

## 3. The case of strong intraband coupling

We focus here on the case characterized by a coupling in the second FS larger than in the first FS. The tuning of the chemical potential is measured by the Lifshitz parameter $z = (\mu - E_{2,L})/\omega_0$ where $E_{2,L}$ is the bottom of the second band and $\omega_0$ is the energy cut-off for the pairing interaction. The cut off interaction is assumed to be the same as the band dispersion ξ. ξ is two times the electronic hopping integral $t_z$ in the direction z that is much smaller that the hopping integrals $t_x$, $t_y$ in the xy direction. The density of states (DOS) in the first 2D Fermi surface is constant while the chemical potential is changed while the ratio of the DOS of the second band changes as shown in panel (a) of Fig. 2.

We assume a Cooper intraband pairing in a first band in the BCS weak coupling regime giving a *first standard BCS condensate,* by fixing the intraband coupling term $c_{11}$=0.23. We consider a case for the second band with an strong attractive pairing interaction $c_{22}$=2.17/$c_{11}$ that could drive the system in a polaronic regime near a band edge.

Where the chemical potential is in regime where the Lifshitz parameter in the range 1<z<-1 a *boson-like condensate* is formed in the second band since all charges in the second band condense. Therefore where the system is tuned to the Lifshitz transition of first type a bipolaron condensate in the strong coupling limit is formed.

We consider now an electronic mechanism that controls the *pair exchange between the two condensates* that is determined by the strength of the interband coupling. The evolution of the critical temperature and of the gap ratios for several values of the interband couplings are shown in the panels (a-c) of Fig. 2.

The minima of $T_c$ are due to the shape anti-resonances below the band edge energy of the second band [panel (a)]. The maximum value of $T_c$ due to the shape resonance is reached near the 3D-2D ETT of the second band, on its 2D side [panel (b)].

We observe that shape resonances in the gap to $T_c$ ratio $\frac{2\Delta_1}{T_c}$ shows clear evidence for the typical asymmetric Fano-lineshape of quantum resonances driven by configuration interaction between a open and closed scattering channels or a quasi-stationary state with a continuum. The antiresonance due to negative interference effects appears in the range between -1<z<0 where $T_c$ has a minimum as it is shown in panel (a) followed by the $T_c$ resonance maximum in the range z≈1.5

At high interband coupling, the $\frac{2\Delta_1}{T_c}$ shows two minima due to the negative interference effect associated with the shape resonance at z=0 and z=-1 and the ratio $\frac{2\Delta_2}{T_c}$ shows an increasing divergence in the mixed Bose-Fermi regime toward z=-1. For intermediate exchange-like interband coupling, we find that there is a critical value $c_{12}>1.52\sqrt{c_{11}c_{22}}$ that separates the strong from the weak interband pairing regime at the band edge. In the weak interband regime $c_{12}<<1.52\sqrt{c_{11}c_{22}}$ the minimum of $\frac{2\Delta_1}{T_c}$ at z≈0 becomes dominant. The maximum of $\frac{2\Delta_1}{T_c}$ moves toward z=0 and at the critical value of the interband pairing [dotted line in panel (c) of Fig. 2] shows a well defined maximum at z=0 and in the range z>1 is well below 3.5, with a typical value around 2, while at high interband coupling it is at the standard BCS value of 3.5.

**4. The case of weak intraband coupling**

Fig. 3 shows the case of weak coupling in the second band where the intraband coupling parameters ratio is fixed at $c_{22}/c_{11}$=0.45. The critical temperature $T_c$ (panels a ), the ratio $\frac{2\Delta_1}{T_c}$ (panel b), and the ratio $\frac{2\Delta_2}{T_c}$ (panel c) are plotted as function of the reduced Lifshitz parameter z. The different curves in each panel represent the cases of different interband pairing strengths ($c_{12}/c_{11}$=-0.68, -0.91, -1.04, -1.59, -2.73). In correspondence of the largest interband pairing strength examined here, $c_{12}$= -2.73, we have obtained critical temperatures as large as 50K, in the range 1<z<2. The antiresonance in the first gap appears when the ratio $\frac{2\Delta_1}{T_c}$ reaches the minimum value, that can be much smaller than standard BCS value ($\frac{2\Delta}{T_c}=3.53$). The antiresonance appears in the range -1<z<0 and moves from -1 to zero decreasing the interband pairing. Moreover the second superconducting gap values results to be nonzero even before the lower band edge $E_{2,L}$ is reached. We notice that the maximum $T_c$ is reached in the BCS regime zone near the type II Lifshitz transition (ETT$_{3D-2D}$).

## 5. Conclusions

The present work provides an interpretation for the properties of superconducting multilayers and indicate a possible roadmap for the discovery of novel HTS like graphene or graphane superlattices. The novel materials should share similarities with the known multigap HTS families where the *shape resonance* is driven by the interband pairing mechanisms.

Here, we have investigated the case of strong intra-band coupling in the second band at the Lifshitz transition (called diboride case) and the case of weak intra-band coupling in the second band at the Lifshitz transition (called pnictide case) like in 11, 1111 and 122 pnictides. In these cases we have changed in our calculation the exchange-like inter-band interaction above and below $c_{12}=1.52\sqrt{c_{11}c_{22}}$ Where $c_{12}>1.52\sqrt{c_{11}c_{22}}$ is clearly the case of iron pnictides that exhibit critical temperatures of the order of 50K mostly driven by interband pairing. The ratios $\frac{2\Delta_1}{T_c}$ and $\frac{2\Delta_2}{T_c}$ show large deviations from the BCS standard value 0.35 and their variation with the chemical potential can be used to identify the superconducting regime for each set of materials.

The antiresonance typical of shape resonance occurs where the Lifshitz energy parameter is zero (i.e. appearing of a new Fermi surface spot) for weak interband exchange pairing and shift to -1 for a very strong interband exchange pairing. The relevant result is that the maximum of the critical temperature appears near the Lifshitz transition of the type "Neck collapsing" in corrugated cylindrical surface as it was well described for a generic heterostucture at atomic limit [16-18] and in previous theoretical works [29] for superlattices of stripes.

**Acknowledgements**

We thank Andrea Perali, Ilya Eremin, Vladimir Kresin and Andrei Shanenko for useful discussions. We gratefully acknowledge partial financial aid from Sapienza University research grant.

REFERENCES

1. J. G. Bednorz and K. A. Müller, Reviews of Modern Physics 60, 585 (1988), URL http://dx.doi.org/10.1103/RevModPhys.60.585.
2. K. A. Muller and A. Bussmann-Holder "Superconductivity in complex systems" Springer-Verlag (Berlin, 2005) ISBN 978-3-540-23124-0
3. A Bianconi, "Symmetry and Heterogeneity in high temperature superconductors" Nato Science Series II Mathematics, Physics and Chemistry vol 214 Springer Dordrecht 2006 The Netherlands ISBN: 10 1-4020-3987-5
4. N. Poccia, A. Ricci, D. Innocenti, and A. Bianconi, International Journal of Molecular Sciences **10**, 2084 (2009), ISSN 1422-0067, URL http://dx.doi.org/10.3390/ijms10052084.
5. G. Bianconi and O. Rotzschke, Physical Review E 82, 036109+ (2010), URL http://dx.doi.org/10.1103/PhysRevE.82.036109.
6. G. Bianconi, Physical Review E 66, 056123+ (2002), URL http://dx.doi.org/10.1103/PhysRevE.66.056123.
7. A. Bianconi, A. Congiu-Castellano, M. De-Santis, P. Rudolf, P. Lagarde, A. M. Flank, and A. Marcelli, Solid State Communications 63, 1009 (1987), ISSN 00381098, URL http://dx.doi.org/10.1016/0038-1098(87)90650-8
8. A. Bianconi M, De Santis, A. Di Cicco, A. M. Flank, A. Fontaine, P. Lagarde, H. K. Yoshida, A. Kotani and A. Marcelli, Physical Review B 38 7196 (1988) URL http://dx.doi.org/10.1103/PhysRevB.38.7196
9. S. Ideta, K. Takashima, M. Hashimoto, T. Yoshida, A. Fujimori, H. Anzai, T. Fujita, Y. Nakashima, A. Ino, M. Arita, et al., Physical Review Letters 104, 227001+ (2010), URL http://dx.doi.org/10.1103/PhysRevLett.104.227001.
10. A. Bianconi, N. L. Saini, A. Lanzara, M. Missori, T. Rossetti, H. Oyanagi, H. Yamaguchi, K. Oka, and T. Ito, Physical Review Letters **76**, 3412 (1996), URL http://dx.doi.org/10.1103/PhysRevLett.76.3412.
11. K. McElroy, D.-H. Lee, J. E. Hoffman, K. M Lang, E. W. Hudson, H. Eisaki, S. Uchida, J. Lee, J.C. Davis, Phys. Rev. Lett. 94, 197005 (2005).
12. Kenjiro K. Gomes, Abhay N. Pasupathy, Aakash Pushp, Shimpei Ono, Yoichi Ando, and Ali Yazdani, Nature 447, 569 (2007).
13. A. Bianconi, N. L. Saini, S. Agrestini, D. Di Castro, and G. Bianconi, International Journal of Modern Physics B **14**, 3342 (2000), URL http://dx.doi.org/doi:10.1142/S0217979200003812.

14. A. Bianconi, S. Agrestini, G. Bianconi, D. Di Castro, and N. L. Saini, Journal of Alloys and Compounds **317-318**, 537 (2001), ISSN 09258388, URL http://dx.doi.org/10.1016/S0925-8388(00)01383-9.
15. M. Fratini, N. Poccia, A. Ricci, G. Campi, M. Burghammer, G. Aeppli, and A. Bianconi, Nature {\bf 466}, 841 (2010), ISSN 0028-0836, URL http://dx.doi.org/10.1038/nature09260.
16. A Bianconi Solid State Communications 89, 933 (1994) ISSN 00381098 URL http://dx.doi.org/10.1016/0038-1098 94 90354-9
17. A Bianconi "High $T_c$ superconductors made by metal heterostructures at the atomic limit" European Patent EP0733271 Sept 25, 1996 (priority date Dec 7, 1993) URL http://v3.espacenet.com/publicationDetails/biblio?DB=EPODOC&adjacent=true&locale=en_EP&FT=D&date=19960925&CC=EP&NR=0733271A1&KC=A1
18. A Bianconi 'Process of increasing the critical temperature Tc of a Bulk Superconductor by Making Metal Heterostructures at the Atomic Limit' US Patent 6265019, July 24, 2001, URL http://patft1.uspto.gov/netacgi/nph-Parser?Sect1=PTO1&Sect2=HITOFF&d=PALL&p=1&u=%2Fnetahtml%2FPTO%2Fsrchnum.htm&r=1&f=G&l=50&s1=6265019.PN.&OS=PN/6265019&RS=PN/6265019
19. A. Bianconi, Solid State Communications 91, 1 (1994), ISSN 00381098, URL http://dx.doi.org/10.1016/0038-1098(94)90831-1.
20. A. Bianconi and M. Missori, Solid State Communications 91, 287 (1994), ISSN 00381098, URL http://dx.doi.org/10.1016/0038-1098(94)90304-2.
21. A. Bianconi, M. Missori, H. Oyanagi, and H. Yamaguchi (SPIE, 1994), vol. 2158, pp. 78-85, edited by: Davor Pavuna, Ivan Bozovic URL http://dx.doi.org/10.1117/12.182701.
22. A. Bianconi, M. Missori, H. Oyanagi, H. Yamaguchi, Y. Nishiara, S. Della Longa EPL (Europhysics Letters) **31**, 411 (1995), ISSN 0295-5075, URL http://dx.doi.org/10.1209/0295-5075/31/7/012.
23. J. T. Devreese and A. S. Alexandrov, Reports on Progress in Physics 72, 066501+ (2009), ISSN 0034-4885, URL http://dx.doi.org/10.1088/0034-4885/72/6/066501.
24. L. Proville and S. Aubry, The European Physical Journal B - Condensed Matter and Complex Systems 15, 405 (2000), URL http://dx.doi.org/10.1007/s100510051142.
25. A. S. Alexandrov and J. Beanland, Physical Review Letters 104, 026401+ (2010), URL http://dx.doi.org/10.1103/PhysRevLett.104.026401.
26. J.M. Blatt and C.J. Thompson Phys. Rev. Lett. 10, 332 (1963).
27. A. Perali, A. Bianconi, A. Lanzara, and N. L. Saini, Solid State Communications 100, 181 (1996), ISSN 00381098, URL http://dx.doi.org/10.1016/0038-1098(96)00373-0

28. A. Perali, A. Valletta, G. Bardeiloni, A. Bianconi, A. Lanzara, and N. L. Saint, Journal of Superconductivity **10**, 355 (1997), ISSN 0896-1107, URL http://dx.doi.org/10.1007/BF02765718.
29. A. Valletta, G. Bardelloni, M. Brunelli, A. Lanzara, A. Bianconi, and N. Saini, Journal of Superconductivity **10**, 383 (1997), ISSN 0896-1107, URL http://dx.doi.org/10.1007/BF02765723.
30. A. Bianconi, A. Valletta, A. Perali, and N. L. Saini, Physica C: Superconductivity 296, 269 (1998), URL http://dx.doi.org/10.1016/S0921-4534(97)01825-X.
31. A. Shanenko and M. D. Croitoru, Physical Review B 73, 012510+ (2006), URL http://dx.doi.org/10.1103/PhysRevB.73.012510.
32. N. Kristoffel, P. Konsin and T. Ord, Rivista Nuovo Cimento 17, 1 (1994)
33. A. Bianconi, D. D. Castro, S. Agrestini, G. Campi, N. L. Saini, A. Saccone, S. D. Negri, and M. Giovannini, Journal of Physics: Condensed Matter **13**, 7383 (2001), ISSN 0953-8984, URL http://arxiv.org/abs/cond-mat/0103211.
34. A. Bianconi, S. Agrestini, and A. Bussmann-Holder, Journal of Superconductivity **17**, 205 (2004), ISSN 0896-1107, URL http://dx.doi.org/10.1023/B:JOSC.0000021214.52321.ab.
35. S. Agrestini, C. Metallo, M. Filippi, L. Simonelli, G. Campi, C. Sanipoli, E. Liarokapis, S. De Negri, M. Giovannini, A. Saccone, et al., Physical Review B **70**, 134514+ (2004), cond-mat/0408095, URL http://dx.doi.org/10.1103/PhysRevB.70.134514.
36. A. Bianconi, Journal of Superconductivity **18**, 25 (2005), ISSN 1557-1939, URL http://dx.doi.org/10.1007/s10948-005-0047-5.
37. A. Bianconi Iranian Journal of Physics Research 6 139-147 (2006)
38. A. Bianconi and M. Filippi in Symmetry and Heterogeneity in high temperature superconductors edited by A. Bianconi (Spinger Dordrecht 2006) Nato Science Series II Mathematics, Physics and Chemistry 214, 21-53 (2006). ISBN-13: 978-1402039881.
39. T. E. Weller, M. Ellerby, S. S. Saxena, R. P. Smith, and N. T. Skipper, Nature Physics **1**, 39 (2005), ISSN 1745-2473, URL http://dx.doi.org/10.1038/nphys0010.
40. A. Ricci, N. Poccia, G. Ciasca, M. Fratini, and A. Bianconi, Journal of Superconductivity and Novel Magnetism **22**, 589 (2009), ISSN 1557-1939, URL http://dx.doi.org/10.1007/s10948-009-0473-x.
41. M. Fratini, R. Caivano, A. Puri, A. Ricci, Z.-A. Ren, X.-L. Dong, J. Yang, W. Lu, Z.-X. Zhao, L. Barba, et al., Superconductor Science and Technology **21**, 092002+ (2008), ISSN 0953-2048, URL http://dx.doi.org/10.1088/0953-2048/21/9/092002.
42. A. Ricci, N. Poccia, B. Joseph, L. Barba, G. Arrighetti, G. Ciasca, J. Q. Yan, R. W. McCallum, T. A. Lograsso, N. D. Zhigadlo, et al., Physical Review B 82, 144507+ (2010), URL http://dx.doi.org/10.1103/PhysRevB.82.144507

43. R. Caivano, M. Fratini, N. Poccia, A. Ricci, A. Puri, Z.-A. Ren, X.-L. Dong, J. Yang, W. Lu, Z.-X. Zhao, et al., Superconductor Science and Technology 22, 014004+ (2009), ISSN 0953-2048, URL http://dx.doi.org/10.1088/0953-2048/22/1/014004
44. V. Chubukov, D. V. Efremov, and I. Eremin, Physical Review B 78, 134512+ (2008), URL http://dx.doi.org/10.1103/PhysRevB.78.134512.
45. V. B. Zabolotnyy, D. V. Evtushinsky, A. A. Kordyuk, D. S. Inosov, A. Koitzsch, A. V. Boris, G. L. Sun, C. T. Lin, M. Knupfer, and B. Büchner, Physica C: Superconductivity 469, 448 (2009), ISSN 09214534, URL http://dx.doi.org/10.1016/j.physc.2009.03.043.
46. C. Liu, T. Kondo, R. M. Fernandes, A. D. Palczewski, E. D. Mun, N. Ni, A. N. Thaler, A. Bostwick, E. Rotenberg, J. Schmalian, et al., Nature Physics 6, 419 (2010), ISSN 1745-2473, URL http://dx.doi.org/10.1038/nphys1656.
47. D. V. Evtushinsky, D. S. Inosov, V. B. Zabolotnyy, M. S. Viazovska, R. Khasanov, A. Amato, H. H. Klauss, H. Luetkens, C. Niedermayer, G. L. Sun, et al., New Journal of Physics 11, 055069+ (2009), ISSN 1367-2630, URL http://dx.doi.org/10.1088/1367-2630/11/5/055069.
48. A. Bianconi, Phys. Stat. Sol. (a) **203**, 2950 (2006), ISSN 18626300, URL http://dx.doi.org/10.1002/pssa.200567003.
49. A. Bianconi, *J. of Physics and Chemistry of Solids* **67**, 566-569 (2006).
50. J. Annett, F. Kusmartsev, and A. Bianconi, Superconductor Science and Technology 22, 010301+ (2009), URL http://dx.doi.org/10.1088/0953-2048/22/1/010301
51. A. Bianconi, N. Poccia, and A. Ricci, Journal of Superconductivity and Novel Magnetism **22**, 526 (2009), ISSN 1557-1939, URL http://dx.doi.org/10.1007/s10948-009-0471-z.
52. D. Innocenti, N. Poccia, A. Ricci, A. Valletta, S. Caprara, A. Perali, and A. Bianconi *Phys. Rev. B* **82** 184528 (2010) `http://dx.doi.org/10.1103/PhysRevB.82.184528`.
53. M. Lifshitz, Zh. Eksp. Teor. Fiz. 38, 1569 (1960) [Sov. Phys. JETP 11, 1130 (I960)].
54. R. Markiewicz, Proc. Int. Conf. on High-Tc Superconductors and Mechanisms and Materials of Superconductivity, Interlaken 1988; Physica C: Superconductivity 153-155, 1181-1182 (1988).
55. A. Varlamov, V. S. Egorov, and A. V. Pantsulaya Adv. Phys. 38 469+ (1989).
56. K. Ikedo, Y. Wakisaka, S. Hirata, K. Takubo, and T. Mizokawa, Journal of the Physical Society of Japan 78, 063707+ (2009), ISSN 0031-9015, URL http://dx.doi.org/10.1143/JPSJ.78.063707.
57. D. Koudela, M. Richter, A. Möbius, K. Koepernik, and H. Eschrig, Physical Review B 74, 214103+ (2006), URL http://dx.doi.org/10.1103/PhysRevB.74.214103.

58. S. Sakai, Y. Motome, and M. Imada, Phys. Rev. Lett. 102, 056404 (2009) and arXiv:1004.2569, (2010) URL http://arxiv.org/abs/1004.2569.
59. Y. Yamaji, T. Misawa, and M. Imada, Journal of Magnetism and Magnetic Materials 310, 838 (2007), ISSN 03048853, URL http://dx.doi.org/10.1016/j.jmmm.2006.10.713.
60. T. Misawa, Y. Yamaji, and M. Imada, Journal of the Physical Society of Japan 77, 093712+ (2008), ISSN 0031-9015, URL http://dx.doi.org/10.1143/JPSJ.77.093712.
61. K. I. Kugel, A. L. Rakhmanov, A. O. Sboychakov, N. Poccia and A. Bianconi Physical Review B 78 165124+ (2008) ISSN 1098-0121 URL http://dx.doi.org/10.1103/PhysRevB.78.165124
62. E. V. L. de Mello and E. S. Caixeiro, *Physical Review B* 70, 224517+, (2004), URL http://dx.doi.org/10.1103/PhysRevB.70.224517.
63. D. Innocenti, A. Ricci, N. Poccia, G. Campi, M. Fratini, and A. Bianconi, Journal of Superconductivity and Novel Magnetism 22, 529 (2009), ISSN 1557-1939 (Print) 1557-1947 (Online), URL http://dx.doi.org/10.1007/s10948-009-0474-9.
64. A. Bianconi, International Journal of Modern Physics B **14**, 3289 (2000), URL http://dx.doi.org/10.1142/S0217979200003769.
65. N. Doiron-Leyraud, et al. Nature 447, 565-568 (2007) URL http://dx.doi.org/10.1038/nature05872.
66. D. LeBoeuf, N. Doiron-Leyraud, B. Vignolle, M. Sutherland, B. J. Ramshaw, J. Levallois, R. Daou, F. Laliberté, O. Cyr-Choinière, J. Chang, et al. (2010), 1009.2078, URL http://arxiv.org/abs/1009.2078.
67. S. Hufner, M. A. Hossain, A. Damascelli, and G. A. Sawatzky, Reports on Progress in Physics 71, 062501+ (2008), ISSN 0034-4885, URL http://dx.doi.org/10.1088/0034-4885/71/6/062501.
68. S. E. Sebastian, N. Harrison, E. Palm, T. P. Murphy, C. H. Mielke, R. Liang, D. A. Bonn, W. N. Hardy, and G. G. Lonzarich, Nature 454, 200 (2008), ISSN 0028-0836, URL http://dx.doi.org/10.1038/nature07095.
69. S. E. Sebastian, N. Harrison, P. A. Goddard, M. M. Altarawneh, C. H. Mielke, R. Liang, D. A. Bonn, W. N. Hardy, O. K. Andersen, and G. G. Lonzarich, Physical Review B 81, 214524+ (2010), URL http://dx.doi.org/10.1103/PhysRevB.81.214524.
70. J. Singleton, et al. 2010 Physical Review Letters 104 086403+ URL http://dx.doi.org/10.1103/PhysRevLett.104.086403.

71. R. Khasanov, S. Strassler, D. Di Castro, T. Masui, S. Miyasaka, S. Tajima, A. B. Holder, and H. Keller, Physical Review Letters 99, 237601+ (2007), URL http://dx.doi.org/10.1103/PhysRevLett.99.237601.

72. M. C. Boyer, W. D. Wise, K. Chatterjee, M. Yi, T. Kondo, T. Takeuchi, H. Ikuta, and E. W. Hudson, Nature Physics 3, 802 (2007), ISSN 1745-2473, URL http://dx.doi.org/10.1038/nphys725.

73. D. L. Feng, A. Damascelli, K. M. Shen, N. Motoyama, D. H. Lu, H. Eisaki, K. Shimizu, Shimoyama, K. Kishio, N. Kaneko, et al., Physical Review Letters 88, 107001+ (2002), URL http://dx.doi.org/10.1103/PhysRevLett.88.107001.

74. A. Chubukov, Physics 3, 54+ (2010), ISSN 1943-2879, URL http://dx.doi.org/10.1103/Physics.3.54.; M. Norman, Physics 3, 86+ (2010) URL http://dx.doi.org/10.1103/Physics.3.86.;

75. N. Kristoffel, P. Rubin and T. Ord Journal of Physics: Conference Series 108 012034+ ISSN 1742-6596 (2008) URL http://dx.doi.org/10.1088/1742-6596/108/1/012034

76. S. G. Ovchinnikov, M. M. Korshunov, E. I. Shneyder Journal of Experimental and Theoretical Physics 109 5 775-785 (2009) URL http://dx.doi.org/10.1134/S1063776109110077

77. M. R. Norman, J. Lin, A. J. Millis Phys. Rev. B 81, 180513+ (2010)

78. S. S. Botelho and C. A. R. S. de Melo, Physical Review B 71, 134507+ (2005), URL http://dx.doi.org/10.1103/PhysRevB.71.134507.

79. D. Innocenti, S. Caprara, N. Poccia, A. Ricci, A. Valletta, and A. Bianconi (2010), Superconductor Science and Technology (in press) preprint http://arxiv.org/abs/1011.4548

80. X. L. Qi, T. L. Hughes, S. Raghu, and S. C. Zhang, Physical Review Letters 102, 187001+ (2009), URL http://dx.doi.org/10.1103/PhysRevLett.102.187001.

81. P. Nikolić, A. A. Burkov, and A. Paramekanti, Physical Review B 81, 012504+ (2010), URL http://dx.doi.org/10.1103/PhysRevB.81.012504.

82. A. Vittorini-Orgeas and A. Bianconi, Journal of Superconductivity and Novel Magnetism 22, 215 (2009), ISSN 1557-1939, URL http://dx.doi.org/10.1007/s10948-008-0433-x.

83. P. Nozieres in A. Griffin, D. W. Snoke, and S. Stringari, *Bose-Einstein condensation* (Cambridge University Press, 1995), p.15. ISBN 9780521589901, URL http://www.worldcat.org/isbn/9780521589901.

84. P. Nozieres J. of Low Temp. Phys.59,195 (1985)

85. A. J. Leggett, in "Modern Trends in the Theory of Condensed Matter", edited by by A. Pekalski and R. Przystawa, Lecture Notes in Physics Vol. 115 (Springer-Verlag, Berlin, 1980), p. 13.

**Figure captions**

**Fig. 1.** Pictorial view of the Fermiology of the system. The evolution of the Fermi surfaces of (FS) the superconductor made of two non-hybridized bands determined by moving the energy separation between the chemical potential and the Lifshitz transitions from panel (a) to panel (d). The first large 2D FS (1) remains nearly constant a 2D cylindrical FS when the chemical potential is changed. The second band goes for the first Lifshitz electronic topological transitions (ETT) panel (b) to the second ETT shown in panel (c). The first ETT occurs moving the chemical potential across the band edge $E_{edge}$ of the second band. The superconductivity phase goes from the single FS in panel (a) with a single condensate to the two FS in panel (b) with two condensates: the first FS (1) has a 2D topology and the second FS (2) has a 3D topology so the two FS's have a different winding number. Where the chemical potential crosses the critical energy $E_{3D-2D}$ the second FS undergoes a 3D-2D ETT shown in panel (c) called disruption of the neck. In fact the second closed 3D FS [panel (b)] becomes the tubular 2D FS in panel (d).

**Fig. 2** The critical temperature in log-scale (panel a) as a function of the Lifshitz parameter $z = (\mu - E_{2,L})/\omega_0$ so that the system crosses the first Lifshitz transition, disappearing of the new Fermi surface spot at z=0 and a second lifshitz transition "disrupting a neck) at z=1. Here we consider the case of strong coupling in the second band (the so-called diboride case or polaronic case) where the intraband Cooper pairing coupling term $c_{22}$ is is about to times large than the Cooper pairing coupling parameter in the first band $\frac{c_{22}}{c_{11}} = 2.17$ where the coupling in the Fermi band is in the standard BCS weak coupling regime $c_{11}$=0.22. The variable ratio of the density of states $N_2/N_1$ (solid black line) is shown in panel (a). The BCS ratios for the first and the second condensates $\frac{2\Delta_1}{T_c}, \frac{2\Delta_2}{T_c}$ as a function of the chemical

potetial are plotted in [panel (b)] and in [panel (c)]. Several values of the interband coupling ratio $c_{12}/c_{11}$ are reported: -6.09 (filled downface triangles); - 4.34 (open diamonds ); -2.24 (filled circles); -1.08 (open squares); -0.43 (filled upwards triangles). The critical temperature reaches 300K, in the range 1<z<2 for the largest value of $c_{12}$. At z=0 we report evidence for the antiresonance due to negative interference effects between different pairing channels that is typical of the quantum nature of shape resonances. $T_c$ reaches the minimum value as shown in panel (a) at the antiresonance for z=0 in the limit of weak interband couping while for a strong interband coupling the mininum due to the antiresonance regimes occurs at z=-1. The gap ratio $\frac{2\Delta_1}{T_c}$ for intermediate values of the interband coupling shows two minima. The second condensate below the band edge for 0<z<-1 is in strong coupling bipolaron bosonic regime far away from the standard Born-Oppenheimer and the Migdal approximations, we see that ratio $\frac{2\Delta_2}{T_c}$ decreases (increases) by approaching the band edge at z=0 for strong (weak) interband interaction. The two types of regimes are separated by the critical value of interband coupling $c_{12} \approx 1.52\sqrt{c_{11}c_{22}}$

**Fig. 3.** The case of weak coupling in the second band (the so called pnictide case) where the intraband coupling parameters ratio is fixed at $c_{22}/c_{11}$=0.45. The critical temperature Tc (panel a), the gap ratio for the first (panel b) and the second (panel c) band are plotted as function of the reduced Lifshitz parameter z. The different curves in each panel represent the cases of different interband pairing strength $c_{12}/c_{11}$: -2.73 (open downface triangles); -1.59 (filled diamonds); -1.04 (open upwards triangles); -0.91 (filled squares) and -0.68 (open circles). For the largest case of interband pairing strength considered here, the critical temperature reaches 300K in the range 1<z <2. The antiresonance regime appears as a minimum in the critical temperature for z=-1 for higher interband repulsive interaction and moves towards z=0 for the smaller interband repulsive interaction. The ratio $\frac{2\Delta_1}{T_c}$ for the first band shows maxima and minima like the critical temperature. On the contrary the ratio $\frac{2\Delta_2}{T_c}$ for the second band in the Bose-like regime shows a decreases (increases) going from threshold to z=0 where it has a minimum (maximum) at for weak (strong) interband repulsive coupling. There are two well defined regimes if the interband coupling is smaller or larger than the critical value $c_{12} \approx 1.52\sqrt{c_{11}c_{22}}$

**Figures**

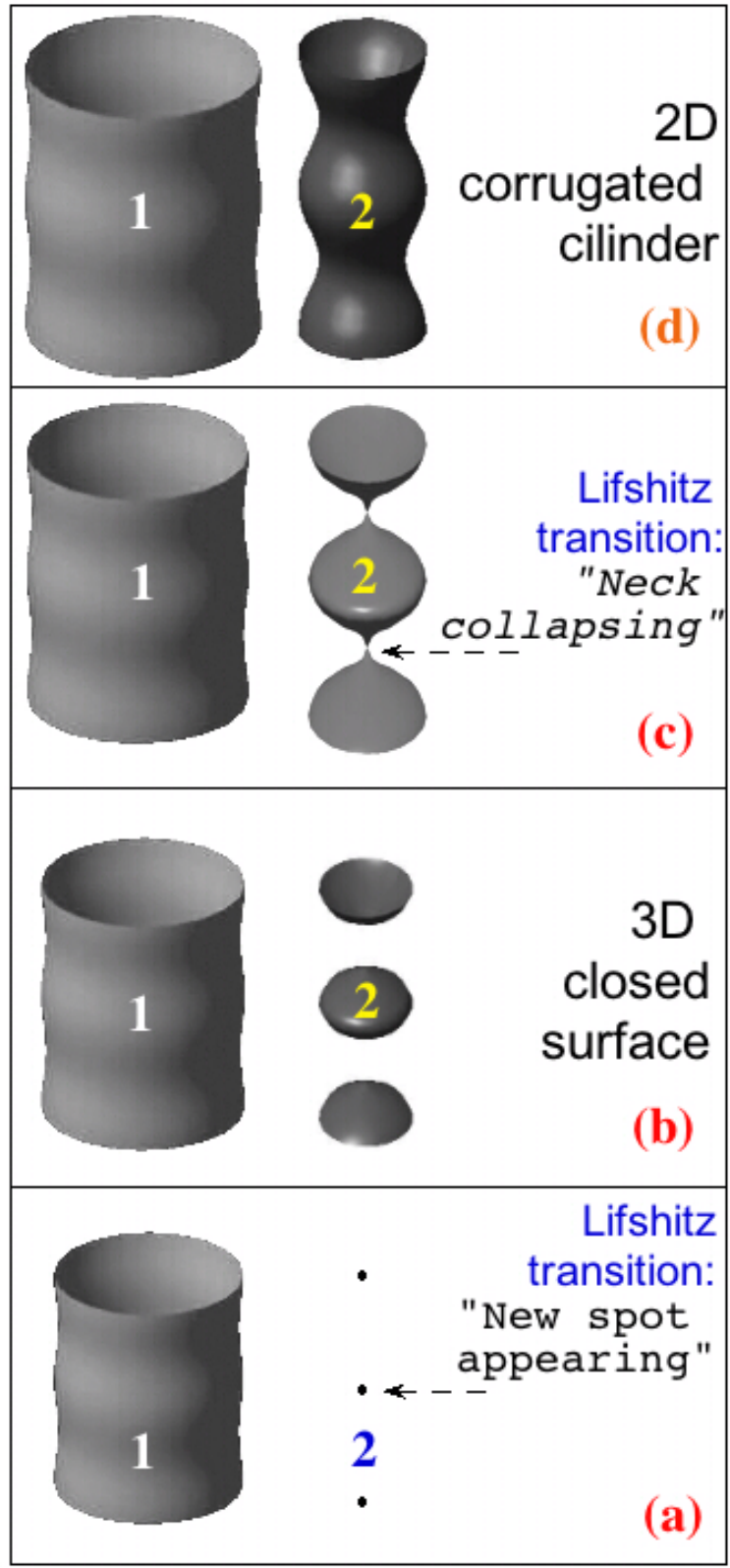

Figure 1.

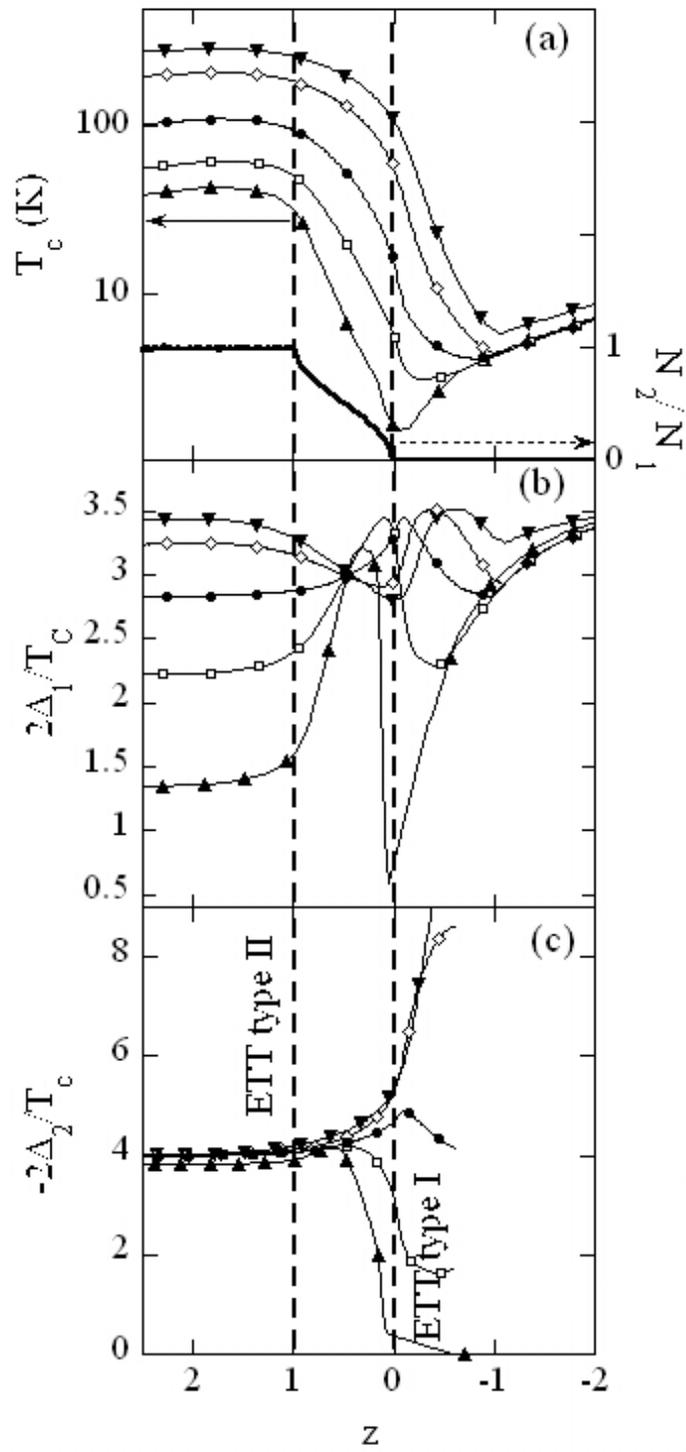

Figure 2.

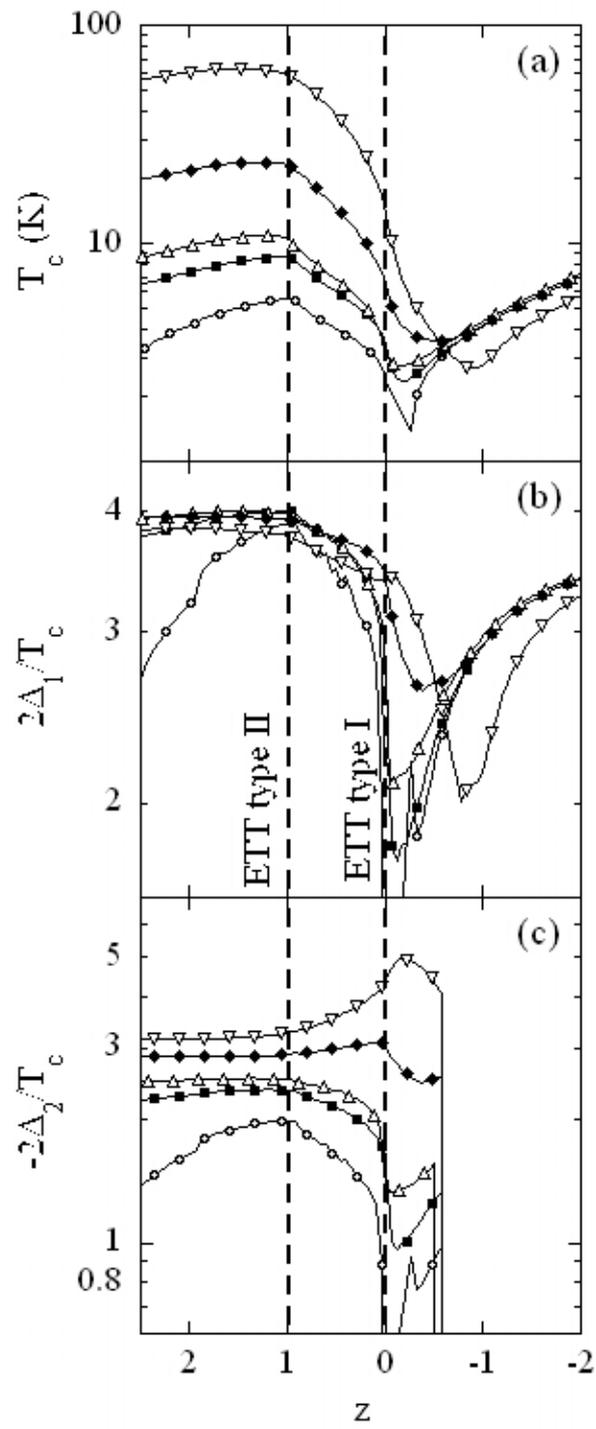

Figure 3.